\documentclass[floatfix,aps,twocolumn,amsmath,amssymb]{revtex4}
\pdfoutput=1

\usepackage{amsfonts, amssymb, amsmath, dsfont}
\usepackage{graphicx}
\usepackage{float}
\usepackage{color}
\usepackage{ulem}
\usepackage[utf8]{inputenc}
\usepackage[T1]{fontenc}

\newcommand{\bm}[1]{\boldsymbol{\mathbf{#1}}}

\newcommand{\Ee}{\operatorname{E}}

\newcommand{\Med}{\operatorname{Med}}
\newcommand{\de}{\operatorname{d}}

\newcommand{\refr}[1]{Ref.~\cite{#1}}

\newcommand{\eq}[1]{Eq.~\eqref{#1}}

\newcommand{\fig}[1]{Fig.~\ref{#1}}

\newcommand{\Fig}[1]{Figure~\ref{#1}}

\newcommand{\beginsupplement}{%
        \setcounter{table}{0}
        \renewcommand{\thetable}{S\arabic{table}}%
        \setcounter{figure}{0}
        \renewcommand{\thefigure}{S\arabic{figure}}%
        \setcounter{section}{0}
        \renewcommand{\thesection}{\arabic{section}}
        \setcounter{equation}{0}
        \renewcommand{\theequation}{S\arabic{equation}}
     }
     
\newcommand{\beginmain}{%
		\renewcommand{\thesection}{\Roman{section}}
     }

\begin{document}

\title{Probing near-field light-matter interactions with single-molecule lifetime imaging}

\author{D. Bouchet}
\author{J. Scholler}
\author{G. Blanquer}
\author{Y. De Wilde}
\author{I. Izeddin}
\altaffiliation{ignacio.izeddin@espci.fr}
\author{V. Krachmalnicoff}
\altaffiliation{valentina.krachmalnicoff@espci.fr}

\affiliation{Institut Langevin, ESPCI Paris, CNRS, PSL University, 1 rue Jussieu, 75005 Paris, France}

\begin{abstract}
Nanophotonics offers a promising range of applications spanning from the development of efficient solar cells to quantum communications and biosensing. However, the ability to efficiently couple fluorescent emitters with nanostructured materials requires to probe light-matter interactions at subwavelength resolution, which remains experimentally challenging. Here, we introduce an approach to perform super-resolved fluorescence lifetime measurements on samples that are densely labelled with photo-activatable fluorescent molecules. The simultaneous measurement of the position and the decay rate of the molecules provides a direct access to the local density of states (LDOS) at the nanoscale. We experimentally demonstrate the performance of the technique by studying the LDOS variations induced in the near field of a silver nanowire, and we show via a Cram\'er-Rao analysis that the proposed experimental setup enables a single-molecule localisation precision of 6~nm.
\end{abstract}

\beginmain

\maketitle

Single fluorescent emitters constitute an excellent probe to access the evanescent near-field of a nanostructure with far-field measurements. Indeed, the advent of super-resolution microscopy in the field of biophotonics has uncapped an unprecedented detail of observation of subcellular structures revealing structural features of tens of nanometres \cite{betzig_imaging_2006,hess_ultra-high_2006,rust_sub-diffraction-limit_2006}, one order of magnitude below the resolution limit imposed by the diffraction of light. While the main super-resolution approaches are based on fluorescence intensity measurements, there exists a strong interest in developing techniques capable of probing lifetime variations at the nanoscale by associating fluorescence lifetime imaging microscopy (FLIM) with subwavelength spatial information. The far-reaching potential of fluorescence lifetime imaging with nanometre resolution is straightforward not only for biological studies \cite{berezin_fluorescence_2010,becker_fluorescence_2012} but also for nanophotonics applications \cite{koenderink_nanophotonics:_2015,fabrizio_roadmap_2016}, as the lifetime of fluorescent emitters is inversely proportional to the LDOS \cite{carminati_electromagnetic_2015}. 

In the last few years, different experimental approaches have been proposed to achieve lifetime measurements at the nanoscale. Super-resolution lifetime imaging was first demonstrated in combination with stimulated emission-depletion (STED) microscopy \cite{auksorius_stimulated_2008}, mostly used for biological applications, and more recently by making use of scanning-probe microscopy to characterise the response of nanostructured plasmonic \cite{frimmer_scanning_2011,krachmalnicoff_towards_2013,beams_nanoscale_2013,schell_scanning_2014,singh_vectorial_2014} or dielectric \cite{bouchet_enhancement_2016} materials to light. Despite the contribution of these methods to nanoscale imaging, a wide-field scheme rather than a scanning approach is essential in order to study dynamic phenomena and to reach molecular resolution. Several groups have recently proposed wide-field approaches to obtain super-resolved LDOS measurements. The association of wide-field localisation with a scanning scheme was used to probe lifetime variations induced by periodic structures \cite{guo_superresolution_2016}. Elegant techniques (although arduous to master) were implemented to measure the lifetime of single quantum dots positioned with microfluidic flow control \cite{ropp_nanoscale_2013,ropp_nanoscale_2015} or using surface-bound motor proteins \cite{gros_parallel_2018}, allowing one to image LDOS variations induced by plasmonic nanostructures. Other methods based on point accumulation for imaging in nanoscale topography (PAINT) \cite{wertz_single-molecule_2015,mack_decoupling_2017} and photo-activated localisation microscopy (PALM) \cite{johlin_super-resolution_2016} need numerical simulations to estimate the LDOS from intensity-based measurements. 

\begin{figure*}[htbp]
\centering
\includegraphics[width=\linewidth]{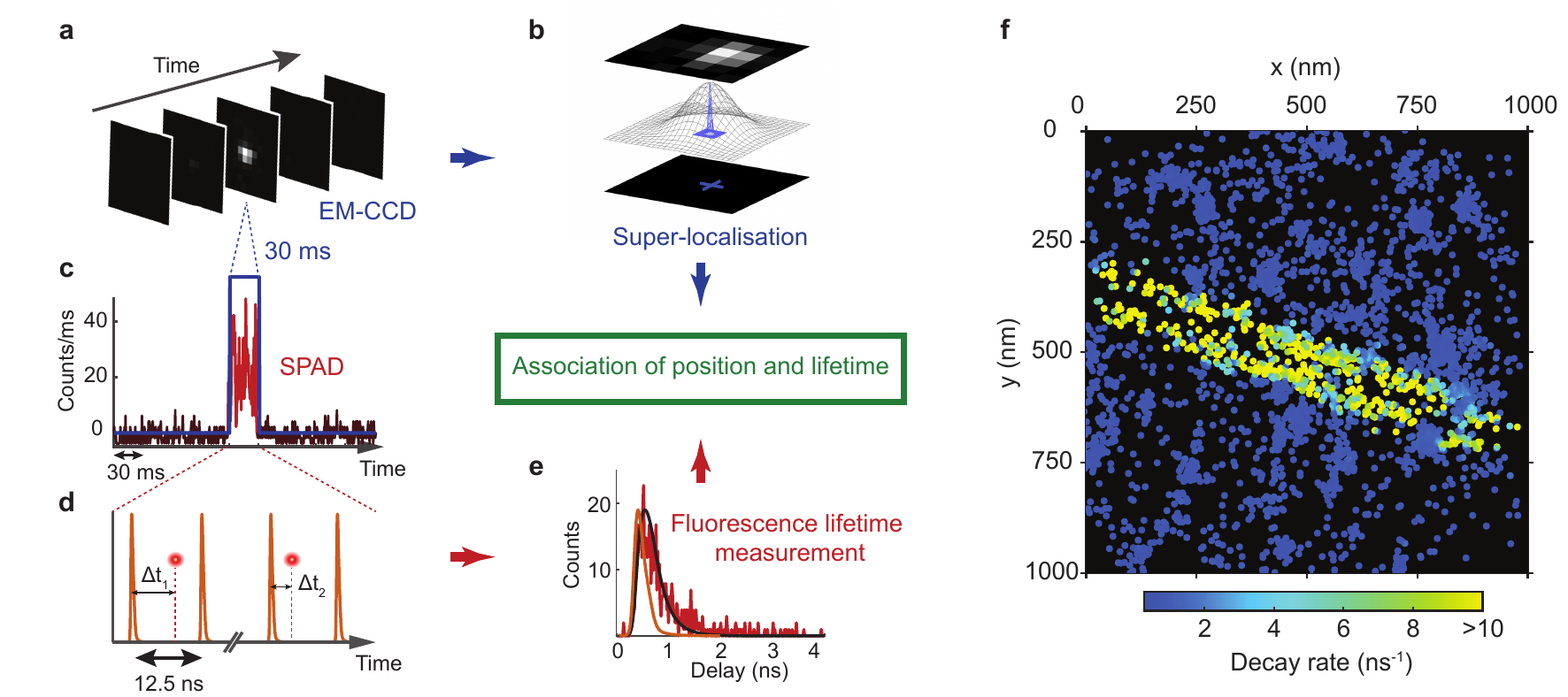} 
\caption{\small Super-resolved LDOS mapping of a silver nanowire. (a)~The EM-CCD camera acquires 31 frames per second, with an exposure time of 30~ms per frame and a field of view of tens of microns. A single fluorescent molecule is detected on the third frame of the sequence shown here. An image cropped around the molecule is shown here for the sake of simplicity. (b)~The position of the molecule is estimated by fitting a two-dimensional Gaussian function to the measured point spread function (PSF). The PSF covers an area of $\sim9$ pixels (pixel size $=160$~nm). (c)~At the same time, the SPAD detects a fluorescence burst from this molecule. (d)~Short laser pulses were used to excite this molecule. For each photon detected during the fluorescence burst, the time difference between excitation and emission can be determined with picosecond precision. (e)~These photons are used to construct a decay histogram. The convolution of the IRF and a decreasing mono-exponential function is then fitted to this histogram in order to estimate the fluorescence decay rate of the molecule. The IRF is shown in orange. (f)~Reconstructed decay rate map. Each dot represents the position of a detected molecule, and its diameter is fixed at 15~nm, which is the typical full width at half maximum (FWHM) of the probability density function followed by the position estimates. If several molecules are detected within the same area, we show their average decay rate on the map.}
\label{f2}
\end{figure*}

In this Letter, we introduce a novel approach that overpasses these limitations and combines lifetime and super-resolved spatial information based on stochastic optical reconstruction microscopy (STORM) \cite{rust_sub-diffraction-limit_2006}, a stochastic imaging technique widely used in biological imaging \cite{huang_breaking_2010}. This method allows to map the lifetime~$\tau$ of stochastically photo-activated single molecules in close vicinity of a densely-labelled nanostructure. It can be readily implemented with a standard microscope and can be applied to biological samples or artificially fabricated nanostructures, either dielectric, metallic, or hybrid metallo/dielectric. Here, we demonstrate the performance of the technique by mapping the LDOS variations induced by a silver nanowire on single molecules located a few nanometers apart. Plasmonic nanowires are an ideal playground to demonstrate the ability of a super-resolved technique to measure light-matter interactions on the nanometer range. They induce strong variations of the lifetime of nearby emitters on the nanometre scale, highlighting the large dynamic range in terms of lifetime modification explorable with our technique. Moreover, due to their geometric simplicity, they enable handleable theoretical studies easily comparable to experimental results.

%\section{Method and experimental results}
%\label{sec:examples}

The sample consists of silver nanowires on a glass coverslip, the whole covered with photo-activatable fluorescent molecules, and is illuminated in wide field with a pulsed laser through an oil immersion objective mounted on an inverted microscope. The studied nanowires have a diameter of $\sim$115~nm and a length of several tens of microns. Their large longitudinal dimension ensures that they weakly radiate to the far-field, therefore strongly limiting the shift in the apparent position of the emitters that has been observed for resonant nanostructures \cite{ropp_nanoscale_2015,wertz_single-molecule_2015,raab_shifting_2017}. The specificity of our method relies on the simultaneous detection of fluorescence photons, through the same microscope objective as the one used for the excitation, on an electron-multiplying charge-coupled device (EM-CCD) camera for super-localisation and on a single-photon avalanche diode (SPAD) coupled to a time-correlated single-photon counting (TCSPC) system for lifetime measurements (see Supplementary Section~1). The EM-CCD camera records wide-field images of the sample with a field of view of tens of micrometers on the sample plane. In contrast, the SPAD, which is a single-channel detector, is conjugated with the center of the camera image via a 50~$\mu$m confocal pinhole and covers an area on the sample plane of $\sim$1~$\mu$m$^2$. By setting the excitation and photoactivation laser power so that no more than one molecule is active at a given time on the area conjugated to the SPAD, the decay rate~$\Gamma=1/\tau$ can be properly estimated for each individual molecule and can be associated to its position.

This approach is illustrated in Fig. \ref{f2}. A single fluorescent molecule is identified on a sequence of wide-field images (Fig. \ref{f2}a), and the position of this molecule is estimated by fitting a two-dimensional Gaussian function to the measured point spread function (Fig. \ref{f2}b). At the same time, the detection of a fluorescent molecule appears as a burst on the signal of the SPAD time trace (Fig. \ref{f2}c). For each SPAD burst, we build the associated decay histogram with a time resolution of 16~ps ( Fig. \ref{f2}d,e). To estimate the decay rate, the convolution of the instrument response function (IRF) and a decreasing mono-exponential function is fitted to the decay histogram. Based on the time correlation between the events detected by the camera and the SPAD (see Supplementary Section~2), we can associate position and decay rate for a large number of photo-activated molecules detected in a single experiment and obtain the super-resolved decay rate map shown in Fig. \ref{f2}f. This map is reconstructed from simultaneous position and decay rate measurements of 3119 molecules, located in a sample region of 1~$\mu$m$^2$ containing one silver nanowire. The typical localisation precision, calculated via a Cramer-Rao lower bound analysis as explained below, is of the order of 6~nm. Spatial variations of the decay rate are observed well below the diffraction limit, demonstrating the ability of the technique to obtain super-resolved LDOS images in a wide-field optical configuration.

%The density of detected molecules is not uniform over the map due to an inhomogeneous labelling (see also Supplementary Section~3). This is not a limitation here since label density does not reflect protein density of the studied structure as is usually required in SMLM for biological applications. The density is here sufficiently high so that the reconstructed image resolution is determined by the localisation precision rather than by sampling considerations. 

\begin{figure}[htbp]
\centering
\includegraphics[width=8cm]{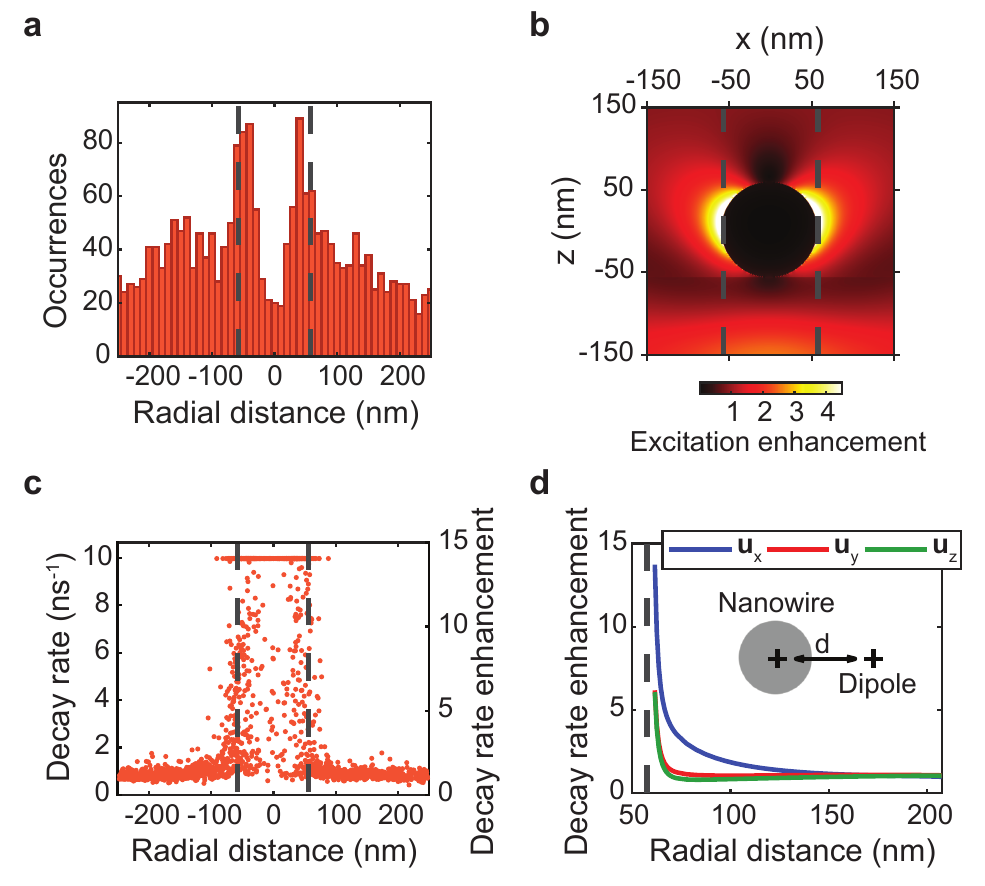} 
\caption{\small (a)~Number of detected molecules as a function of the distance to the nanowire axis. (b)~Time-averaged intensity of the excitation field in the vicinity of the silver nanowire calculated from the results of a FDTD simulation. On these figures, dashed lines represent the estimated position of the nanowire edges. (c)~Distribution of decay rate versus distance to the wire axis. The highest decay rate that can be measured, limited by the IRF of the setup, is  $10$~ns$^{-1}$. (d)~Decay rate enhancement as a function of the distance to the nanowire for the three orientations of the dipole moment. The inset shows a cross-section of the system numerically studied.
}
\label{f3}
\end{figure}

A unique insight as allowed by this new approach is revealed by the study of the density of detected molecules along the center of the nanowire axis (see also Supplementary Section~3). Fig. \ref{f3}a shows that, on average, twice as many molecules are detected for a distance to the nanowire axis $d = \pm50$~nm than for $d = 0$~nm. Indeed, the interaction between the excitation field and the nanowire results in a non-uniform excitation intensity distribution, as shown in Fig. \ref{f3}b by finite-difference time-domain (FDTD) simulations (see Supplementary Section~4). A local enhancement of the excitation intensity is observed on the sides of the nanowire, with a lateral extension of about 20~nm, as well as an extinction of the excitation intensity on the top of the wire. Therefore, the molecules located in the higher excitation intensity regions have a larger probability to be detected, supporting the observed variations of the density of detected molecules. Furthermore, the image is formed by a two-dimensional projection of fluorescent events around a cylindrical nano-object. This also affects the apparent density of detected molecules.

In order to get a deeper insight into the observed decay rate variations, we further studied the dependence of the decay rate on the distance $d$ to the nanowire axis (Fig. \ref{f3}c). Molecules detected far from the nanowire axis ($d > 200$~nm) show an average value of the decay rate of 0.68~ns$^{-1}$ with a standard deviation of 0.17~ns$^{-1}$. In contrast, the decay rate is higher than 10~ns$^{-1}$ for many molecules detected at distances $d < 60$~nm from the nanowire axis. This leads to a decay rate enhancement of a factor 15, only limited by the IRF of the setup (see Supplementary Section~5). This measurement confirms that molecules with the largest decay rates are those attached to the nanowire or in its closest vicinity. We further numerically simulated the enhancement of the decay rate induced by the presence of the nanowire for three orthogonal dipole moment orientations (Fig. \ref{f3}d). Experimental and numerical results are in good qualitative agreement, supporting the validity of the experimental technique. Different dipole moment orientations can explain the lifetime dispersion observed in the vicinity of the nanowire. %\emph{j'ai enleve \cite{cruz_quantitative_2016}}

%Despite the lack of information about the orientation of the molecules \cite{cruz_quantitative_2016}, we can see that different dipole moment orientations can explain the lifetime dispersion observed in the vicinity of the nanowire.

%\section{Discussion}

\begin{figure}[htbp]
\centering
\includegraphics[width=\linewidth]{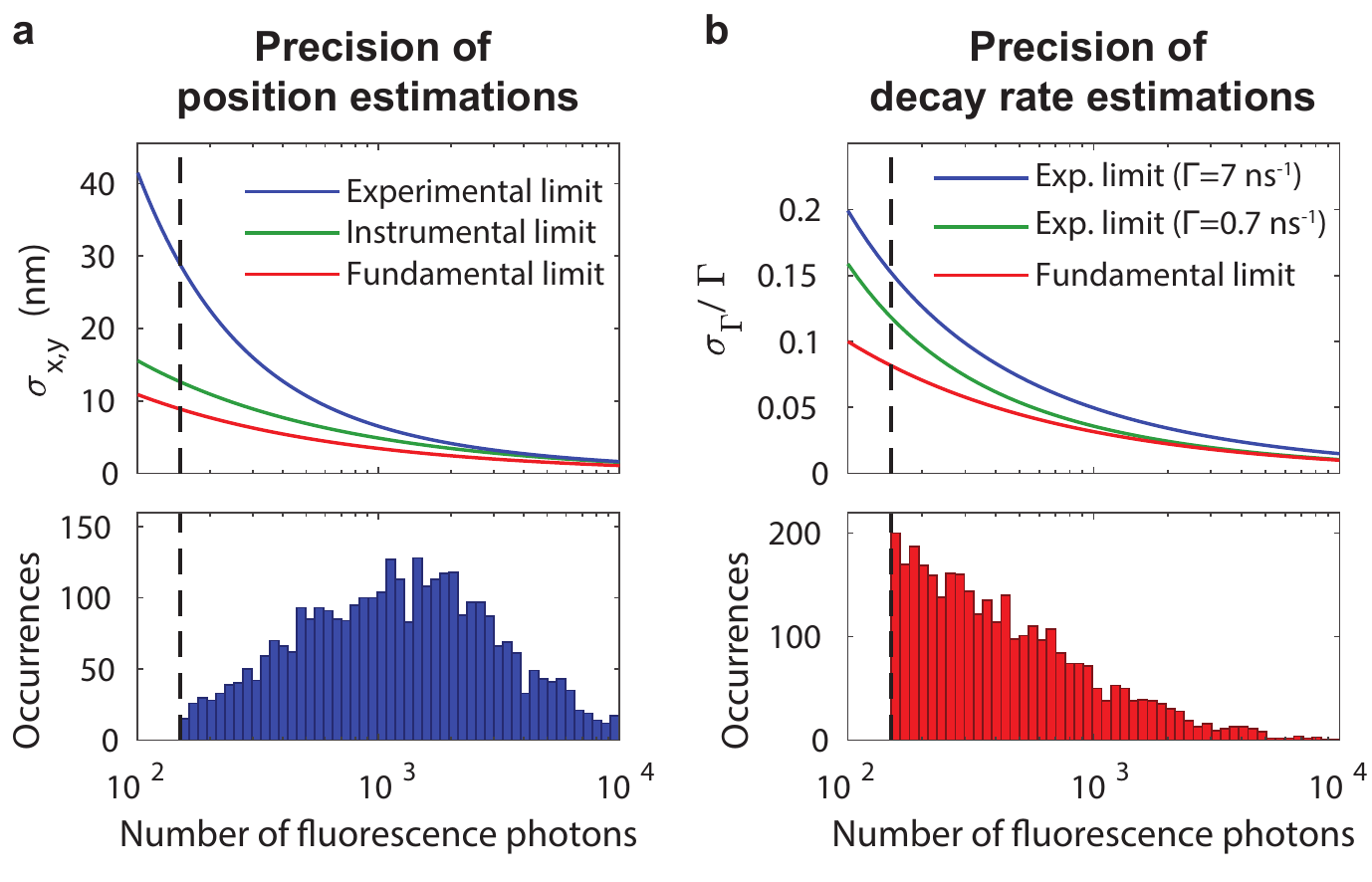} 
\caption{\small (a)~Top: Cram\'{e}r-Rao bound on the standard error on the position estimates as a function of the number of fluorescence photons detected by the camera. Bottom: Distribution of the number of fluorescence photons detected by the camera from the single molecules. (b)~Top: Cram\'{e}r-Rao bound on the standard error on the decay rate estimates as a function of the number of fluorescence photons detected by the SPAD. Bottom: Distribution of the number of fluorescence photons detected by the SPAD from the single molecules. On these figures, dashed lines represent the threshold condition $N>150$ used for data analysis.}
\label{f4}
\end{figure}

The performance of the proposed method ultimately relies on the precision at which we can estimate both the position and the decay rate of the detected single fluorophores. We can assess a lower bound on these parameters by calculating the Cram\'{e}r-Rao lower bound \cite{kay_fundamentals_1993} on the standard error of the position and lifetime estimators, respectively noted $\sigma_{x,y}$ and $\sigma_{\Gamma}$ (see Supplementary Sections~6 and~7). Such analysis is standard in localisation microscopy to assess the localisation precision \cite{deschout_precisely_2014,chao_fisher_2016}. Fig. \ref{f4}a shows the dependence of $\sigma_{x,y}$ upon the number of fluorescence photons detected by the EM-CCD camera. The fundamental limit (red curve) is set by the shot noise and the finite pixel size as sources of error on the measurement. The instrumental limit (green curve) also accounts for the readout noise of the camera and the noise introduced by the electron multiplying process. The actual limit of our experiment (blue curve) is calculated by considering  additional sources of noise such as substrate luminescence. The number of fluorescence photons experimentally detected by the camera from each molecule ranges from 150 to more than $10^4$ fluorescence photons with a median value of 1228 photons (\ref{f4}a, bottom). With this value the Cram\'{e}r-Rao bound for position estimations is 6~nm.

A similar analysis can also be performed for lifetime estimations \cite{kollner_how_1992,bouchet_Fisher_2018}. The fundamental limit on the relative standard error of decay rate estimators $\sigma_{\Gamma}/\Gamma$ is simply given by $1/\sqrt{N}$, where $N$ is the number of detected photons (Fig. \ref{f4}b, red curve). We calculated the Cram\'{e}r-Rao bound for $\Gamma = 0.7$~ns$^{-1}$ (molecules on glass) and $\Gamma = 7$~ns$^{-1}$ (molecules close to the nanowire), which corresponds to a lifetime of 140~ps, comparable to the FWHM of the IRF (240~ps). As expected, $\sigma_{\Gamma}/\Gamma$ deviates from the fundamental limit when the number of measured fluorescence photons is smaller than 1000 due to the influence of background noise. In the experiment, the median value of detected photons is 367~photons (Fig. \ref{f4}b, bottom). For this value, $\sigma_{\Gamma}/\Gamma$ ranges from 8\% to 10\% depending on the value of $\Gamma$. %Note that the typical numbers of collected photons is lower for the SPAD than for the camera. This is due to the use of a pinhole on the optical path leading to the SPAD and the difference in the quantum efficiency of two detectors ($\sim$0.5 for the SPAD and $\sim$0.9 for the EM-CCD). These two effects explain the differences in the distribution of the number of photons detected by each detector. 
%
%\section{Conclusion}
The Cram\'{e}r-Rao analysis thus demonstrates that the proposed experimental setup enables state-of-the-art measurements of light-matter interactions with a localisation precision of 6~nm together with a relative error of 10\% for lifetime estimations. Future prospects will include accessing the axial position of the detected molecules with the implementation of three-dimensional localisation methods \cite{hajj_accessing_2014,chizhik_metal-induced_2014,bourg_direct_2015,isbaner_axial_2018}. The technique can notably be adapted to perform three-dimensional imaging using metal-induced energy transfer, as suggested by a recent article reporting three-dimensional localisation for sparsely distributed single molecules~\cite{karedla_three-dimensional_2018}. Additionally, by taking advantage of SPAD arrays constituted of several independent channels \cite{cuccato_complete_2013}, a field of view of tens of micrometres in the sample plane could be reached, opening a wide range of interesting opportunities for imaging and sensing applications. The acquisition time could also be reduced by actively optimising the number of molecules simultaneously photo-activated on the region conjugated to the SPAD. Multiple simultaneous detections could be treated with an improved detection, fitting, and reconstruction algorithm.

The readiness of the technique to be implemented with a standard microscope suggests a great potential to rapidly expand into a wide variety of applications, ranging from nanophotonics and plasmonics to biophotonics. Topical applications in nanophotonics include the direct characterisation of samples presenting rich LDOS patterns and strongly-confined electromagnetic fields, with concrete perspectives for the study of light localisation in strongly scattering media \cite{sapienza_long-tail_2011,riboli_engineering_2014}. The technique is not constrained to the fluorphores used in the present realisation, but can be extended to photoactivatable fluorophores of different wavelengths and to DNA-PAINT for an \textit{a priori} knowledge of the fluorophore position. Thanks to these extensions, it will be possible to characterise the resonant and non-resonant behavior of a nanostructure and to tackle the mislocalisation of resonant fluorophores with fluorescence lifetime measurements. In the field of biophotonics, wide-field FLIM images with nanometre resolution will allow to probe local dynamic phenomena in living cells. We also foresee that, by associating our approach with techniques based on F\"orster resonance energy transfer (FRET), \textit{in cellulo} nanoscale imaging of molecule-molecule interactions will soon become within reach.

%This method also opens various possibilities in the field of biophotonics. Indeed, wide-field FLIM images with nanometre resolution opens new avenues to probe local fluctuations in living cells. We also foresee that, by associating our approach with techniques based on Förster resonance energy transfer (FRET), nanoscale imaging of molecule-molecule interactions will soon become within reach. % Future prospects will include accessing the axial position of the detected molecules with the implementation of three-dimensional localisation methods \cite{hajj_accessing_2014,chizhik_metal-induced_2014,bourg_direct_2015}. Additionally, by taking advantage of SPAD arrays constituted of several independent channels \cite{cuccato_complete_2013}, a field of view of tens of micrometres in the sample plane could be reached, opening a wide range of interesting opportunities for imaging and sensing applications.

\section*{Funding information}
This work was supported by LABEX WIFI (Laboratory of Excellence ANR-10-LABX-24) within the French Program Investments for the Future under reference ANR-10- IDEX-0001-02 PSL*, by the Programme Emergences 2015 of the City of Paris, and by ANR-17-CE09-0006 SimpleLife.

\section*{Acknowledgments}
The authors thank S.~Bidault for helping in sample preparation, A.~C.~Boccara for sharing his insights about the manuscript and I.~Rech, A.~Gulinatti and A.~Giudice for providing the PMD-R detector.

\bibliographystyle{apsrev_no_url}

%\begin{small}
%\bibliography{references}
%\end{small}

%%%%%%%%%%%%%%%%%%%%%%%%%%%%%%%%%%%%%%%%%%%%%%%%%%%%%%%%%%%%%%%%%%%%%%%%%%%%%%%%

\onecolumngrid
\pagebreak
\beginsupplement
\begin{center}
\textbf{\large Probing near-field light-matter interactions with single-molecule lifetime imaging: supplementary material}

\bigskip
D. Bouchet, J. Scholler, G. Blanquer, Y. De Wilde, I. Izeddin, and V. Krachmalnicoff\\
\textit{\small Institut Langevin, ESPCI Paris, CNRS, PSL University, 1 rue Jussieu, 75005 Paris, France}
\end{center}
\vspace{1cm}
\twocolumngrid

\section{Experimental setup}

\subsection{Sample preparation}

To prepare the sample, we spin-coat a dilute solution of silver nanowires in isopropyl alcohol on a glass coverslip. A microfluidic chamber is then prepared as follows \cite{lermusiaux_widefield_2015}: we cover the sample with a ring made of parafilm, we place two micro-pipettes on opposite sides of the parafilm ring and we cover them with another glass coverslip before heating the sample up to 70$^{\circ}$C in order to melt the parafilm. We let the microfluidic chamber cool down for a few minutes before using the micro-pipettes to inject biotin diluted in a phosphate-buffered saline (PBS) solution at a concentration of 1~g/L. We leave this solution incubate for 2~hours. Then, we inject streptavidin-conjugated fluorescent molecules (Alexa 647) diluted in a PBS solution at a concentration of 0.005~g/L, and we leave this new solution incubate for 2~hours. We add a PBS solution containing a few polystyrene fluorescent beads 100~nm in diameter (Red FluoSpheres, ThermoFisher Scientific) which we use as fiducial markers, and we then fill the chamber with an oxygen-reducing buffer \cite{heilemann_subdiffraction-resolution_2008}. This buffer is prepared according to the following protocol \cite{van_de_linde_direct_2011}: we use a PBS solution in which we dilute dextrose (100~mg/mL), cysteamine (3.86~mg/mL), glucose oxidase (0.5~mg/mL) and catalase (1.18~$\mu$L of an aqueous solution concentrated at 20-50 mg/mL).

\subsection{Optical setup}

\begin{figure}[ht]
\centering
\includegraphics[width=8cm]{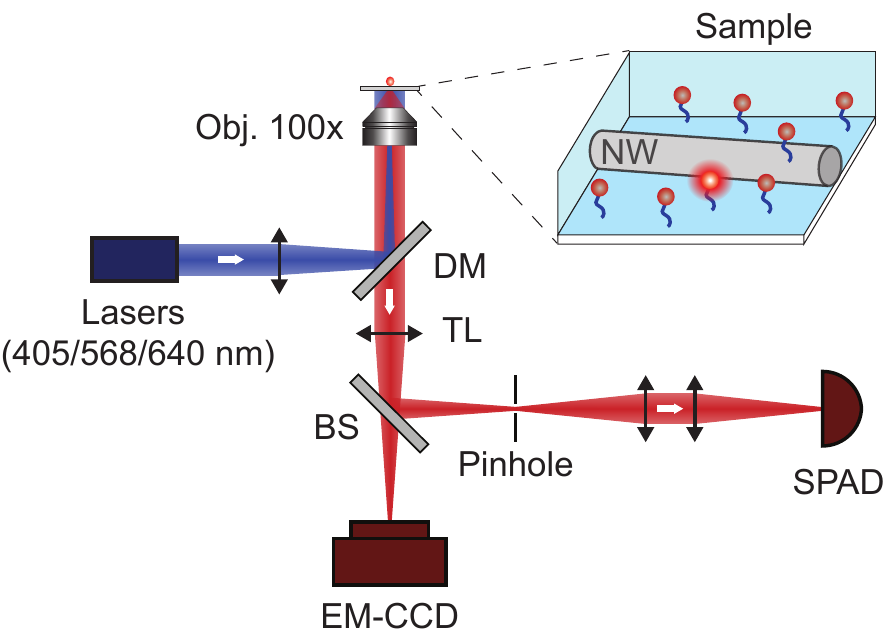} 
\caption{\small Optical setup. The excitation laser ($\lambda=640$~nm), together with the photo-activation laser ($\lambda=405$~nm) and the laser used for sample stabilisation ($\lambda=568$~nm), illuminate the sample via a high numerical aperture oil objective (NA=1.4). A lens (f=300~mm) is located on the excitation path to ensure wide-field illumination. Fluorescence from the sample is filtered with a dichroic mirror (DM) and passes through a tube lens (TL). A 50:50 beamsplitter (BS) splits the light towards an EM-CCD camera and a SPAD. The sample under study contains photo-activated single molecules in the near-field of a silver nanowire (NW).}
\label{f1}
\end{figure}

Before the experiment, we select an area on the sample in which a silver nanowire can be identified by basic transmission imaging, thus ensuring that only one nanowire is present in the detection volume. Then, we place the area of interest in the middle of the field of view of the camera by using a piezoelectric stage (PXY 200SG,
Piezosystem Jena). Photo-activatable molecules (Alexa Fluor 647) are excited by a pulsed laser diode emitting at $\lambda=640$~nm (LDH Series P-C-640B, PicoQuant) at a repetition rate of 80~MHz. The intensity incident on the sample averaged over a repetition period is 10~$\mu$W/$\mu$m$^2$. %As each pulse is characterised by a full width at half maximum (FWHM) of 240~ps, the effective intensity incident on the sample is approximately 500~$\mu$W/$\mu$m$^2$. 
The laser polarisation is set perpendicular to the nanowire axis in order to minimise the backscattering of the laser light by the nanowire. The molecules are photo-activated with a laser diode emitting at $\lambda=405$~nm (LDH Series P-C-640B, Picoquant). During the acquisition, the density of activatable molecules decreases in time since several molecules are photobleached by the excitation laser. To compensate for this effect, we progressively turn on the photo-activation laser, with an average intensity on the sample up to 50 nW/$\mu$m$^2$. A third laser (Fianium SC450) filtered at $\lambda=568$~nm is required for the excitation of fiducial markers that are used for real-time drift correction. 
%This laser is a supercontinuum laser (Fianium SC450) filtered at $\lambda=568$~nm by an excitation filter (LL01-568, Semrock). 
These three lasers illuminate the sample through an oil immersion objective (UPLSAPO 100XO, NA=1.4, Olympus) mounted on an inverted microscope (\fig{f1}).
% and characterised by a $\times 100$ magnification and 1.4 numerical aperture. Before the objective, a lens characterised by a focal length of 300~mm is used to obtain a 
Wide-field illumination over an area of approximately 200~$\mu$m$^2$ is achieved by placing a lens ($f=300$~mm) before the objective.
Fluorescence from the sample is then collected by the objective and filtered by a dichroic mirror 
%(ZT405/488/561/647rpc, Chroma) 
as well as two long-pass filters.
%(ZET405/488/561/640m, Chroma and FF01-446/523/600/677, Semrock). 
Then, a 50:50 beamsplitter splits the signal towards two paths. On the first path, fluorescence photons are directed towards an EM-CCD camera (iXon~897, Andor). On the second path, a SPAD (PDM-R, Micro Photon Devices~\cite{gulinatti_new_2012}) is connected to a time-correlated single-photon counting (TCSPC) system (HydraHarp400, Picoquant). %This setup allows the determination of decay rates up to 10~ns$^{-1}$, corresponding to fluorescence lifetimes of 100~ps. On this path, we use a 50~$\mu$m confocal pinhole to conjugate the SPAD to a small area on the sample plane ($\sim$800~nm in diameter). 

\subsection{Drift correction}

To determine and correct the drift in the sample plane, we estimate the position of a fiducial marker from the wide-field images acquired by the camera and we use a feedback loop to maintain the marker at a fixed position. Every 5~s, the drift is estimated by fitting a two-dimensional Gaussian function to the image of the marker. A feedback signal is then applied on the piezoelectric stage (PXY 200SG,
Piezosystem Jena) controlling the in-plane position of the sample in order to compensate for the drift. %We operate the piezoelectric positioning system in open-loop mode to eliminate additional noise introduced by the sensor and the driver stabilisation system.

In order to estimate the drift of the sample in the axial direction with respect to the focal plane, we analyse images of the fiducial marker accumulated over several seconds. The defocus-correction system is based on a real-time maximisation of the power spectral density of the measured images, and the axial position of the objective with respect to the sample is corrected in real time with a piezoelectric positioning system (MIPOS 20SG, Piezosystem Jena) located between the objective and the microscope turret.

\section{Position and decay rate association}

\subsection{Position estimations}

The EM-CCD camera acquires 31 frames per second with an acquisition time of 30 ms per frame. The full sequence of wide-field images saved by the camera (over a subset of 13$\times$13~pixels, pixel size $=160$~nm) is imported by ImageJ \cite{schneider_nih_2012} and the positions of the photo-activated molecules are estimated using ThunderSTORM~\cite{ovesny_thunderstorm:_2014}. First of all, each frame is filtered using a wavelet filter, as proposed by Izeddin \textit{et al.}~\cite{izeddin_wavelet_2012}. For each frame, approximate localisation of the molecules is then performed by applying a threshold that depends on the signal-to-noise ratio of the camera data. For this acquisition, we set it to 2.7 times the standard deviation of the intensity values obtained in the filtered image. Finally, sub-pixel localisation of the molecules is performed by fitting a two-dimensional Gaussian function to the data using the weighted least squares method on a restricted domain around the molecule (7$\times$7~pixels). %As an illustration, we present here a frame in which a single molecule can be identified (\fig{ch5_example_fig3}a); a two-dimensional Gaussian function is fitted to the measured PSF in order to estimate the position of the molecule (\fig{ch5_example_fig3}b). 
As some molecules can be identified over consecutive frames, we perform a merging of the data acquired by the camera if the estimated distance between successive detections is less than 40~nm. Then, the position of the molecule is determined by using the average value of the positions estimated from the different frames. Using this strategy, we obtain approximately 24,000 different detections for the whole experiment. This number is limited by the weak activation power required to ensure that no more than one molecule is typically active at a single time on the area conjugated to the SPAD.

%\begin{figure}[ht]
%\begin{center}
%\includegraphics[scale=0.95]{ch5_example_fig3}
%\hspace{0.5cm}
%\includegraphics[scale=0.95]{ch5_example_fig4}
%\end{center}
%\caption{(a)~Detection of a single molecule by the EM-CCD camera. (b)~Gaussian function fitted to the measured PSF. Black lines represent the estimated coordinates of the molecule.}
%\label{ch5_example_fig3}
%\end{figure} 

\subsection{Decay rate estimations}

In addition to EM-CCD images, we also record the arrival time of each photon detected by the SPAD. To deal with the large size of the resulting file ($\sim$15~GB), the 10-hour-long acquisition is split into several sequences of approximately 50~minutes. Then, we compute the number of detected photons as a function of time with a resolution of 500~$\mu$s. The intensity of background noise associated with this signal usually decreases during the experiment due to a decreasing number of activated molecules in the periphery of the detection area. Hence, the intensity time trace is Fourier filtered in order to remove low frequency components associated with temporal fluctuations longer than 30~s. Then, we consider that a molecule is potentially detected for each burst surpassing a given threshold that depends on the signal-to-noise ratio of the SPAD data. For each 50-minutes-long sequence, we set it to 2.6 times the standard deviation of the filtered signal. If another burst occurs within the typical blinking time scale (20~ms), it is attributed to the same molecule. %As an illustration, we present here a small part of the intensity time trace in which a burst can be identified (\fig{ch5_example_fig2}a). 
In total, we identify approximately 14,000 events over the 10-hour-long acquisition. This value is small in comparison to the number of detections obtained from camera data. Indeed, the area of the sample conjugated to the SPAD (see \fig{ch5_hole_fig1A}b) is smaller than the area over which the localisation is performed ($\sim 1100\times1100$~nm).
For each SPAD event, we build the associated decay histogram with a resolution of 16~ps in order to estimate the decay rate. To do so, the contribution of background noise is estimated by using close-by time intervals in which no burst can be identified. Then, the convolution of the instrument response function (IRF) and a decreasing mono-exponential function is fitted to the decay histogram using the least-squares method. The value of the decay rate is set to 10~ns$^{-1}$ if the fit yields a value higher than this limit. Indeed, the IRF of the setup is characterised by a FWHM of approximately 240~ps (corresponding to 4~ns$^{-1}$) and we consider that estimates above 10~ns$^{-1}$ are not meaningful even after the deconvolution process. While sample heterogeneities could induce multi-exponential decays, the small number of photons detected by the SPAD from each molecule does not allow to resolve different lifetimes. For this reason, we restrict the analysis to a mono-exponential decay, which would therefore correspond to an average over different decays.

%\begin{figure}[ht]
%\begin{center}
%\includegraphics[scale=0.95]{ch5_example_fig2}
%\hspace{0.5cm}
%\includegraphics[scale=0.95]{ch5_example_fig1}
%\end{center}
%\caption{(a) Identification of a burst in the intensity time trace -- this burst corresponds to the detection shown in \fig{ch5_example_fig3}. (b) Estimation of the decay rate from the decay histogram. The estimated decay rate is here 4.7~ns$^{-1}$.}
%\label{ch5_example_fig2}
%\end{figure} 

%\section{Temporal and spatial correlations}
%
%The detections identified from camera data and the bursts identified from SPAD data are strongly time-correlated. We precisely characterise the spatio-temporal correlation between SPAD events and camera detections in order to associate the position of a large number of molecules with their decay rate. For simplicity, we use the subscript $i$ when referring to a detection identified using camera data and the subscript $j$ when referring to an event identified using SPAD data.

\subsection{Temporal and spatial correlations}

\begin{figure}[ht]
\begin{center}
\includegraphics[scale=0.85]{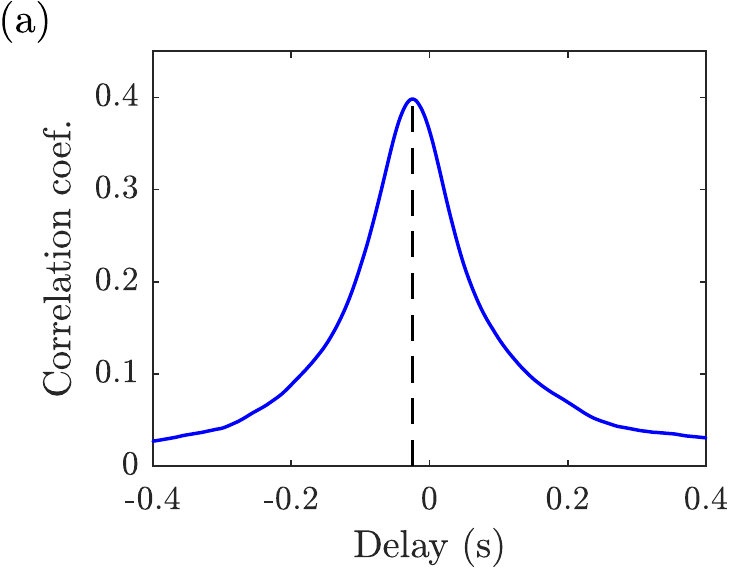}

\vspace{0.6cm}
\hspace{0.4cm}\includegraphics[scale=0.85]{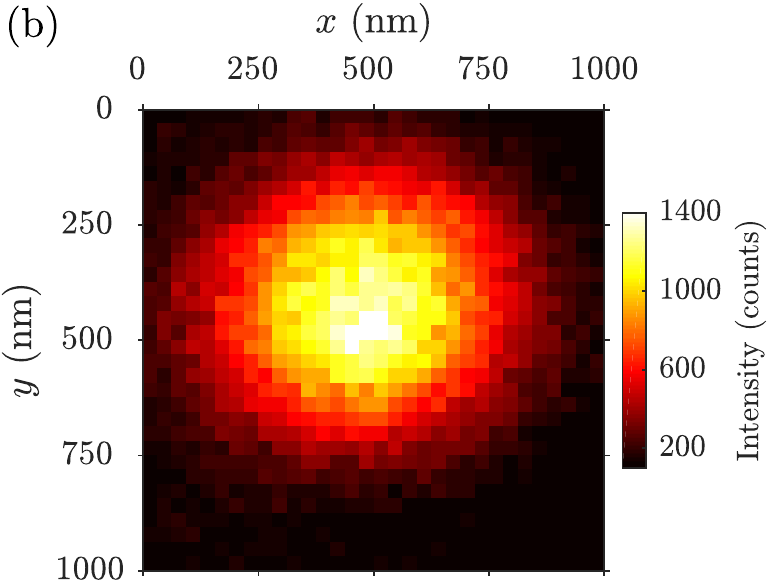}
\end{center}
\caption{(a)~Correlation coefficient calculated from binary representations of the SPAD events and the camera detections. A dashed line represents the estimated time offset between the two channels. (b)~Measured response of the SPAD while scanning a fluorescent bead in the sample plane.}
\label{ch5_hole_fig1A}
\end{figure}

It is important to keep in mind that the SPAD does not include information about the position of the molecules. We therefore need to ensure that the lifetime information provided by the SPAD is properly associated with the position of the molecules provided by the EM-CCD camera. At the beginning of the experiment, the acquisition of both camera and SPAD data is started by using an in-house software, and we can expect a time offset of several milliseconds between the two different channels. In order to precisely determine this time offset, we build two binary representations respectively associated with the SPAD events and the camera detections (1 for a SPAD event or a camera detection, 0 otherwise). We then calculate the time correlation of these binary representations with a resolution of 500~$\mu$s, as shown in \fig{ch5_hole_fig1A}a for a typical sequence of 50~minutes. The maximum of this correlation coefficient gives an accurate estimate of the time offset between the camera and the SPAD. This delay is typically around 20~ms, which is consistent with the data acquisition procedure. Note that the correlation coefficient does not reach unity but is typically between 0.3 and 0.5. Indeed, the conditions required for the detection of a molecule by the camera and by the SPAD are different. In comparison to the SPAD, the camera is characterised by a larger field of view and a larger quantum efficiency. However, its lower temporal resolution makes the identification process less efficient for molecules characterised by fast temporal fluctuations. Hence we can expect some molecules to be detected by only one of the two detectors, resulting in a value smaller than unity for the maximum of the correlation coefficient.

In order to characterise the spatial correlation between SPAD events and camera detections, we must identify the pixels of the camera that are conjugated to the area of the sample seen by the SPAD. Hence, we measure the response of the SPAD by scanning a fluorescent bead with a diameter of 100~nm over a large area in the sample plane. \Fig{ch5_hole_fig1A}b shows the number of photons detected by the SPAD as a function of the bead position. The FWHM value of the measured profile is of the order of 500~nm, as expected from the diameter of the confocal pinhole (50~$\mu$m) and the magnification of the optical system ($\times 100$). This response can be modelled by a function $h(x,y)$ which is the convolution of a 500~nm gate and a two-dimensional Gaussian function.

%\begin{figure}[ht]
%\begin{center}
%\includegraphics[scale=0.95]{time_correlation_fig1}
%\end{center}
%\caption{}
%\label{ch5_time_correlation_fig1}
%\end{figure} 

\subsection{Association conditions}

Once the time offset between the camera and the SPAD is estimated and compensated, we can quantify the time overlap between a camera detection and a SPAD event. To do so, we simply calculate the ratio of the time overlap $\Delta t_{ij}$ to the time interval $\Delta t_{j}$ corresponding to the SPAD event. The camera detection and the SPAD event are likely to be associated to the same molecule whenever this ratio is close to unity. We can then associate position and decay rate in the following situations:
\begin{itemize}
\item In 77\% of the cases, the association between position and decay rate is straightforward. In such cases, only one camera detection is identified in the emission time $\Delta t_j$ corresponding to a SPAD event. In addition, this SPAD event is the only one identified in the emission time $\Delta t_i$ corresponding to the camera detection. Therefore, the camera detection $i$ and the SPAD event $j$ can be associated.
\item In 18\% of the cases, several camera detections at different positions are identified in $\Delta t_j$. In such cases, we can estimate the number of photons to be detected by the SPAD from a given camera detection. Let $x_i$ and $y_i$ be the coordinates in the sample plane corresponding to a detection and $N_i$ the number of fluorescence photons measured by the camera, we can simply assume that the number of photons to be detected by the SPAD is proportional to $N_i \, h(x_i,y_i)$. 
%One can compare the quantum efficiencies of the camera ($\sim 0.9$) and the SPAD ($\sim 0.5$) to obtain an estimate of the number of photons to be detected by the SPAD. This is however not required for a comparison between different detections identified from the camera data. In such cases, we evaluate the likelihood of each detection to be the one corresponding to the SPAD event, based on the number of fluorescence photons measured by the camera, the distance to the centre of the detector and the temporal overlap between the SPAD event and the considered frame. 
An association condition can thus be set on the base of the value taken by $T_{ij} = N_i \, h(x_i,y_i) \Delta t_{ij}/\Delta t_j$.
After the identification of the detection $k$ on the camera associated with the maximum value of $T_{ij}$, we consider that the association between position and decay rate can be performed only if $T_{kj}> \alpha_{a} \, \sum_{i=1}^n T_{ij}$ where $n$ is the number of camera detections in $\Delta t_j$ and $\alpha_a$ is a threshold characterising the association condition. If $\alpha_a$ is low, camera detections are more frequently associated to SPAD events. However, this increases the number of cases in which the measured decay histograms are the sum of different decay histograms that cannot be properly separated by a post-processing analysis. As a trade-off, we use $\alpha_a=80$\% in the experiment.
\item In 5\% of the cases, several SPAD events are identified in $\Delta t_i$. Then, if the difference between these decay rates is smaller than 30\%, we merge the SPAD events and we calculate the average decay rate. Otherwise, we evaluate the likelihood of each event to be the one corresponding to the camera detection, based on the number of fluorescence photons measured by the SPAD. To do so, we identify the event $k$ associated with the highest number of photons $N_k$ and we perform the association between position and decay rate only if $N_k> \alpha_{a} \, \sum_{i=1}^n N_i$ where $N_i$ is the number of photons associated with the overlapping SPAD events and $\alpha_a$ is the threshold previously mentioned ($\alpha_a=80$\%).
\end{itemize}

\paragraph{Post-process filtering}

Two additional conditions are required in order to correctly perform the association between position and decay rate. For each molecule, at least 150~fluorescence photons must be detected on each detector. Moreover, the standard deviation of the Gaussian function fitted to the camera data must be smaller than 190~nm. These two conditions avoid the occurrence of false detections that would be due to noise. Using this procedure, we associate the position of 3,581 camera detections with their decay rate. We then perform post-processing filtering to account for the few remaining loopholes of the procedure. To do so, we compare each decay rate to the decay rate of the 10~closest detections. On average, this corresponds to a distance of 19~nm between the detection and its neighbours. Then, we perform an outlier identification based on the median absolute deviation (MAD). A decay rate $\Gamma$ is rejected if the decay rates $\Gamma_k$ of the closest neighbours satisfy the following condition:
\begin{equation}
| \Gamma - \Med(\Gamma_k) | > \alpha_r \; \Med \left[ \cfrac{| \Gamma_k - \Med(\Gamma_k) |}{0.675} \right] \; ,
\label{ch5_equation1}
\end{equation}
where $\Med$ is the median operator and $\alpha_r$ is a rejection threshold. The factor 0.675 is used so that MAD and standard deviation are approximately equal for large normal samples~\cite{maronna_dispersion_2006}. It should be noted that no outlier identification is performed if more than 50\% of the neighbours have a decay rate equal to the upper limit previously mentioned (10~ns$^{-1}$) since the right-hand side of \eq{ch5_equation1} equals zero in this case. With the approach expressed by \eq{ch5_equation1}, using a small threshold $\alpha_r$ allows the identification of many outliers but may also identify actual detections as outliers. As a trade-off, we use $\alpha_r=5$ resulting in the identification of 6\% of outliers. By removing them, the number of actual detections reduces to 3,352.

\section{Density and intensity maps}

\begin{figure}[ht]
\begin{center}
\includegraphics[scale=0.85]{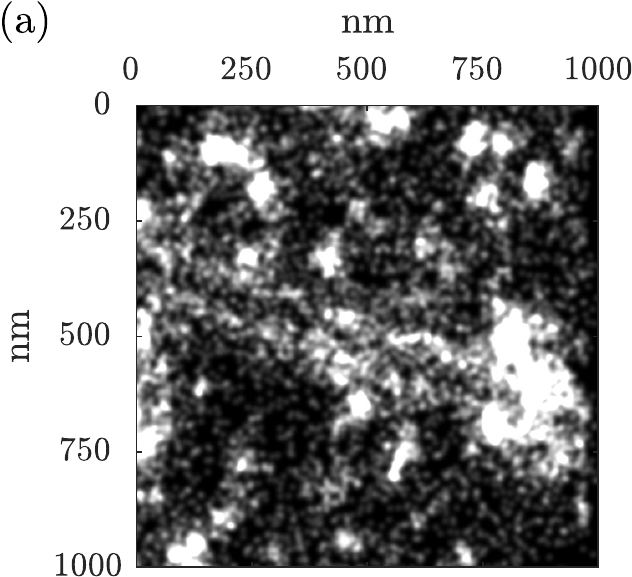}

\vspace{0.6cm}
\hspace{0.6cm}\includegraphics[scale=0.85]{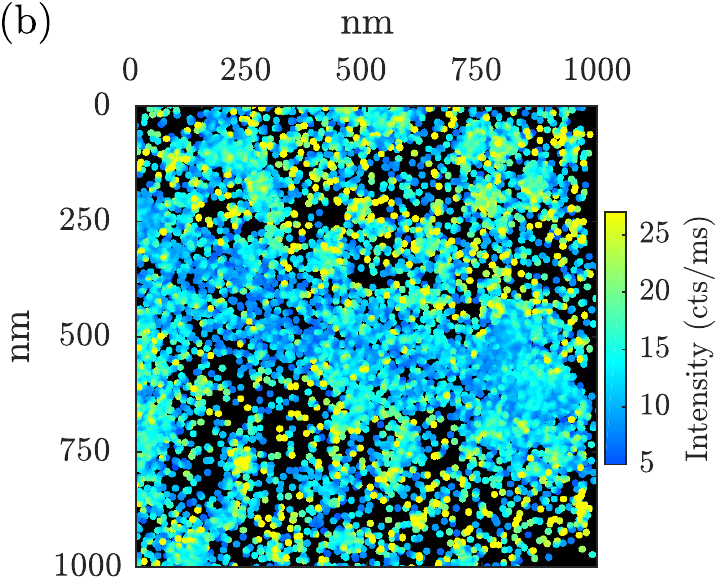}
\end{center}
\caption{(a)~Density and~(b) intensity maps reconstructed from the 14,546 molecules detected by the EM-CCD camera.}
\label{maps}
\end{figure}

From data acquired by the EM-CCD camera, we can render a density map of the detected molecules (\fig{maps}a), as for a usual single-molecule localisation-based super-resolution image reconstruction. In \fig{maps}a, we observe strong density fluctuations due to an inhomogeneous labelling of our sample. However, note that for the purpose of obtaining a map of the LDOS, inhomogeneous labeling is not a limitation given a high enough spatial sampling, which underlines the robustness of our fluorescence lifetime measuring technique. It is important to underline that, in the image reconstruction in \fig{maps}a, the strong density differences renders an image where black regions do not necessarily represent a lack of detections. In the case of biological applications, the labelling is specific to the protein of interest and thus density fluctuations represent structural changes of the sample which is not the case in our LDOS nanocartography.

Additionally, we can also reconstruct a color map coding the measured fluorescence intensity for each detection (\fig{maps}b). Note that, if several molecules are detected within the same area, we plot the average intensity. While the density of detected molecules is higher along the sides of the nanowire than on the substrate, we observe that the collected intensity is lower for the molecules on the nanowire. %This can be explained because, on the one hand, we measure the 2D projection of the 3D object which is the nanowire; on the other hand, 
Indeed, although the excitation field is larger for the molecules on the sides of the nanowire, their radiative quantum yield is reduced due to coupling to non-radiative modes (surface plasmon modes and quenching). 
%This is the result of two competing effects. Indeed, the excitation intensity is larger for these molecules, but the radiative quantum yield is reduced due to coupling to non-radiative modes (surface plasmon modes and quenching).

\section{Numerical simulations}

Simulations are performed using the FDTD simulation software MEEP \cite{oskooi_meep:_2010}. The relative permittivity of silver is modelled with a Lorentz–Drude model, the relative permittivity of the buffer solution is set to 1.77 and the relative permittivity of glass is set to 2.25. In order to estimate the influence of the excitation field on the observed density variations, we model the system in two dimensions, with a mesh resolution of 0.5~nm. The nanowire, located on a glass substrate, is illuminated by a plane wave at $\lambda=640$~nm polarised perpendicularly to the nanowire, as in the experiment. In this configuration, a two-dimensional simulation gives the exact solution due to the invariance of the structure and the source along the longitudinal dimension. In contrast, in order to study the decay rate enhancement due to the nanowire, we model the system in three dimensions, with a mesh resolution of 1~nm. As the effect of the substrate on the decay rate is small due to the low contrast between the relative permittivities of the buffer solution and the glass coverslip, we perform the simulations without the substrate to limit the computational time. In each simulation, the emitter is modelled as an electric dipole source that generates a Gaussian pulse at $\lambda=670$~nm, and the decay rate is estimated from the value of the electric field at the source position. We assume that the intrinsic quantum yield of Alexa Fluor 647 dyes is 0.33, as specified by the provider, in order to calculate the total decay rate enhancement.

\begin{figure*}[ht]
\centering
\includegraphics[scale=0.74]{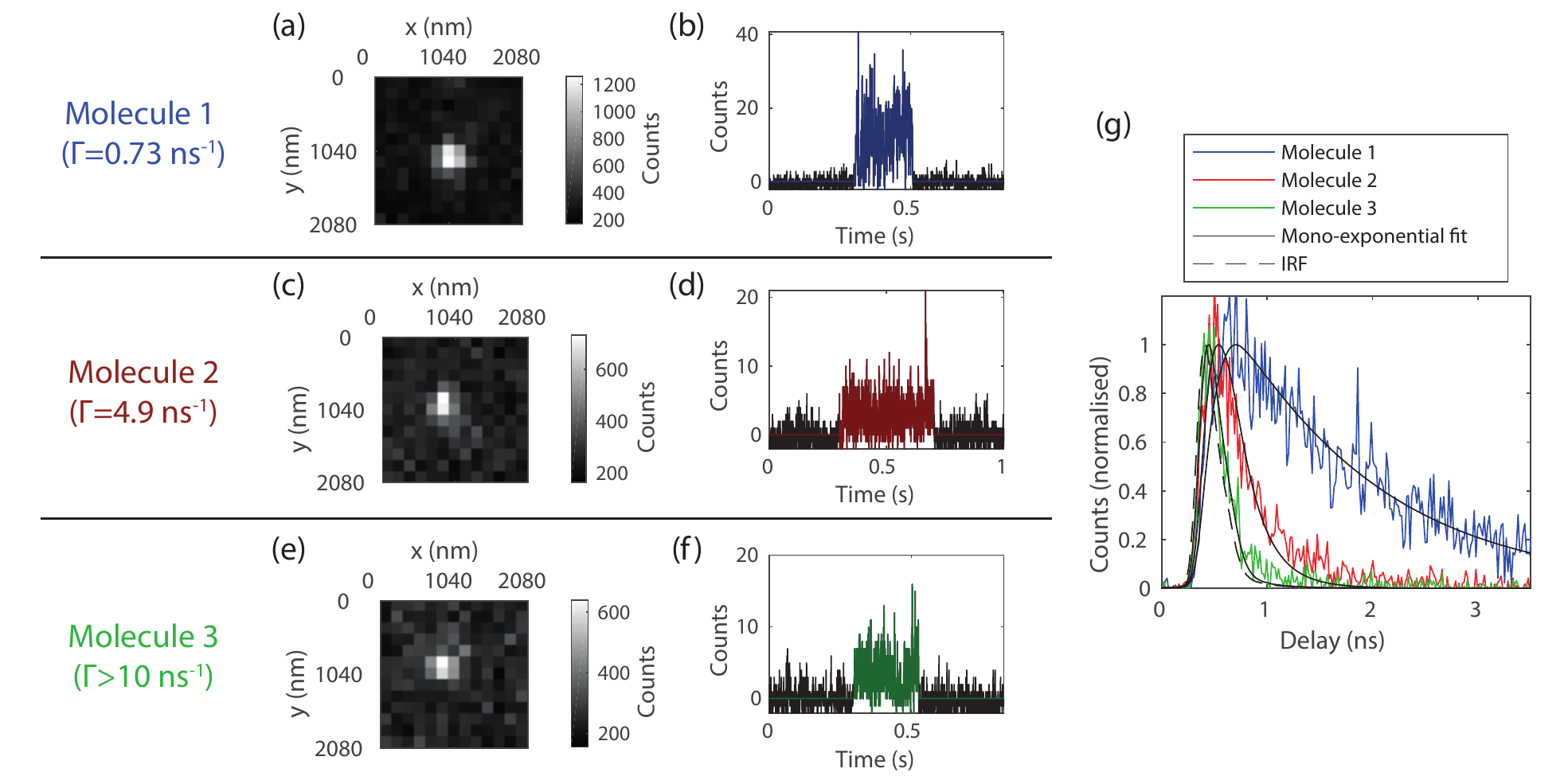}
\caption{Camera images and signal measured by the SPAD for a molecule far from the nanowire [sub-figures (a) and (b)], for a molecule close to the nanowire with $\Gamma/\Gamma_0\sim$7 [sub-figures (c) and (d)], and for a molecule close to the nanowire with $\Gamma/\Gamma_0>$15 [sub-figures (e) and (f)]. The associated decay histograms are shown in sub-figure~(g).}
\label{lifetime3}
\end{figure*}

\section{Decay histograms of single molecules}

In this section, we show decay histograms for different molecules far from the nanowire and in its close vicinity, providing clear evidence of the decay rate enhancement. \Fig{lifetime3}~(a) to~(f) shows the signal measured by the camera and by the SPAD during the experiment for three molecules characterised by different decay rate. \Fig{lifetime3}~(g) shows the associated decay histograms, together with mono-exponential fits. While the decay rate of the molecule far from the nanowire is not enhanced (molecule~1), the decay rate of the two molecules in the close vicinity of the nanowire show a strong decay rate enhancement (molecules~2 and~3). For the third molecule considered, the decay rate cannot be resolved by the current experimental setup as the decay histogram and the IRF are superimposed.

\section{Cramér-Rao analysis: position estimations}

To estimate the Cram\'{e}r-Rao lower bound on the standard error of position estimators $\sigma_{\bar{x},\bar{y}}$, we follow the approach described in \cite{chao_fisher_2016_2}. The data acquired by the EM-CCD camera are modeled using the Airy function to describe the fluorescence signal, as well as a uniform background noise originating from the luminescence of the substrate. Then, we consider that the probability density function describing the number of photoelectrons per pixel is given by the convolution of the amplified signal and the Gaussian readout noise. The information matrix is calculated from this probability density function, and numerically inverted in order to compute the Cram\'er-Rao bound.

\paragraph{Point spread function}

We consider the simple situation in which a far-field microscope is used to collect the photons emitted by a single molecule located in the object plane. We assume that the 2-dimensional probability density function (PDF) describing the intensity distribution in the image plane can be expressed from the coordinates in the image plane noted $(x',y')$ and the coordinates of the molecule in the object plane noted $(x_0,y_0)$ as follows:
\begin{equation}
q(x',y') = \cfrac{J_1^2 \left( \cfrac{2 \pi \mathrm{NA} \sqrt{(x'-Mx_0)^2+(y'-My_0)^2}}{M \lambda_0} \right)}{\pi \left[(x'-Mx_0)^2+(y'-My_0)^2 \right]} \; ,
\label{ch6_equation20}
\end{equation}
where $J_1$ is the first-order Bessel function of the first kind, NA is the numerical aperture of the objective, $M$ is the magnification and $\lambda_0$ is the free-space emission wavelength. The expectation of each data item -- that is, the expectation of the value measured on each pixel by the camera -- is then expressed as follows:
\begin{equation}
\begin{split}
f_i =& N \int\limits_{(x',y') \in pixel} q(x',y') \de x' \de y' \\ &+ N_b \int\limits_{(x',y') \in pixel} q_b(x',y') \de x' \de y' \; ,
\label{ch6_equation7}
\end{split}
\end{equation}
where $N$ is the total number of photons emitted by the molecule and detected by the camera and $N_b$ is the number of photons due to background noise which follows a PDF noted $q_b(x',y')$. In \eq{ch6_equation7}, the integration is performed over the area that defines the considered pixel. Note that a dedicated study of the point spread function in our geometry could  improve the prediction of position of the dipoles along the nanowire \cite{su_visualization_2015}.

\paragraph{EM-CCD data model}

We can now derive a functional form for the likelihood function that describes the number of events measured on each pixel by the camera. Assuming that fluorescence photons detected by the camera are statistically independent, the number of photons impinging on each pixel during a given time interval follows a Poisson distribution of expectation $f_i$. If we do not consider the additional noise arising from the detection process, the PDF associated with the observation of $X$ photoelectrons on a given pixel is
\begin{equation}
p_i^{p}(X; \boldsymbol{\theta}) = \frac{f_i^{X}}{X!} \, e^{-f_i} \; ,
\label{ch6_equation8}
\end{equation}
where $\boldsymbol{\theta}$ are the parameters that must be estimated from the data (here, the parameters are the coordinates of the molecule). This sets the fundamental limit achievable by a perfect camera. However, the multiplication register of an EM-CCD camera enhances the number of generated photoelectrons in order to beat the readout noise of the camera, and the PDF followed by the number of photoelectrons generated by the process depends on the gain $g$. As shown in \refr{chao_fisher_2012}, this PDF noted $p_i^e(X; \boldsymbol{\theta})$ can be approximated, for large gain values, by
\begin{small}
\begin{equation}
p_i^e(X; \boldsymbol{\theta}) = 
\begin{cases}
e^{-f_i} ,  & \text{for } X=0  , \\
 \cfrac{e^{(-X/g-f_i) } \, \sqrt{\cfrac{f_i X}{g}} \; I_1 \left( 2 \sqrt{\cfrac{f_i X}{g}} \right)}{X} ,  & \text{for } X>0 ,
 \end{cases}
\label{ch6_equation6}
\end{equation}
\end{small}
where $I_1$ is the first-order modified Bessel function of the first kind. In addition, the readout process induces a Gaussian noise on each pixel characterised by an expectation $\eta_g$ and a standard deviation $\sigma_g$. This Gaussian noise can be described by the following PDF:
\begin{equation}
p^g(X; \boldsymbol{\theta}) = \frac{1}{\sigma_g \sqrt{2 \pi}} \exp \left(-\frac{(X-\eta_g)^2}{2 \sigma_g^2} \right) \; .
\label{ch6_equation9}
\end{equation}
The PDF describing the readout noise of the camera is the same for all the pixels. Therefore, we can consider that the PDF describing the number of photoelectrons per pixel for a real EM-CCD camera is given by
\begin{equation}
p_i(X; \boldsymbol{\theta}) = \left[ p_i^e(X; \boldsymbol{\theta}) \right] \ast \left[ p^g(X; \boldsymbol{\theta}) \right] \; ,
\label{ch6_equation10}
\end{equation}
where the asterisk ($\ast$) represents the convolution product. Then, the information matrix can be numerically evaluated from its general expression given by~\cite{kay_fundamentals_1993_2}
\begin{equation}
\begin{split}
\left[\bm{\mathcal{I}}(\boldsymbol{\theta})\right]_{jk} =& \sum_{i=1}^n \Ee \left[ \frac{1}{[p_i(X ; \boldsymbol{\theta})]^2} \right. \\
& \left. \times \left( \frac{\partial p_i(X ; \boldsymbol{\theta}) }{\partial \theta_j} \right) \left( \frac{\partial p_i(X ; \boldsymbol{\theta}) }{\partial \theta_k} \right) \right] \; .
\end{split}
\label{ch6_equation4}
\end{equation}

\paragraph{Cramér-Rao bound}
After having experimentally measured the value of the parameters involved in the model, we can compute the Cramér-Rao bound on the variance of position estimators in order to evaluate a lower bound on the standard error $\sigma_{x,y}$ on the position estimates performed using one frame. Assuming that there is no preferred direction in space -- this is not exactly true because of the shape of the pixels, but is a good approximation for squared pixels -- the Cramér-Rao inequality reads
\begin{equation}
\sigma_{x,y} \geq \sqrt{\frac{1}{\mathcal{I}_{xx}}} = \sqrt{\frac{1}{\mathcal{I}_{yy}}} \; .
\end{equation}
In the experiment, a molecule is typically detected on two successive frames. Its position is then estimated by the mean of the individual estimates, so that the standard error on the resulting position estimate is $\sigma_{\bar{x},\bar{y}}=\sigma_{x,y}/\sqrt{2}$. Different situations can then be compared: the fundamental is calculated using \eq{ch6_equation8} with $N_b=0$, the instrumental limit is calculated using \eq{ch6_equation10} with $N_b=0$, and the experimental limit is calculated using \eq{ch6_equation10} with the value of $N_b$ measured in the experiment.

\section{Cramér-Rao analysis: decay rate estimations}

To estimate the Cram\'{e}r-Rao lower bound on the relative standard error of decay rate estimators $\sigma_{\Gamma}/\Gamma$, we adopt a similar approach, described in \cite{bouchet_fisher_2018_2}. In order to estimate the Cram\'{e}r-Rao bound for our experiment, we model the fluorescence decay by the convolution of the IRF and an exponential distribution. After proper inclusion of time-dependent background noise in the model, it can then be considered that each point of the decay histogram follows a Poisson distribution. Thus, we can compute the information matrix from this distribution and numerically invert it in order to obtain the Cram\'er-Rao bound.

\paragraph{SPAD data model}

By modelling a molecule by a two-level system, the PDF that describes the photon emission time $t$ is given by an exponential distribution. Since the agreement between experimental data and the mono-exponential model is satisfactory, we consider here that this model is relevant. Then, the PDF followed by the photon detection time measured by the experimental system is
\begin{equation}
q(t) = q_{irf}(t) \ast \left[ \Gamma e^{- \Gamma t} \right] \; ,
\label{ch6_equation15}
\end{equation}
where $q_{irf}(t)$ is the PDF describing the IRF of the setup. From this expression, we can find the expectation of each data item; that is, the expectation of each data point of the decay histogram. We obtain
\begin{equation}
f_i = N \sum_{l=0}^{+ \infty} \int\limits_{t_i+lT}^{t_{i+1}+lT} q(t) \de t +N_b \int\limits_{t_i}^{t_{i+1}} q_b(t) \de t \; ,
\label{ch6_equation14}
\end{equation}
where $N$ is the number of photons emitted by the molecule and detected by the system, $N_b$ is the number of detected photons due to background noise which follows a PDF noted $q_b(t)$, and $T$ is the repetition period of the laser. If the fluorescence lifetime of the molecule is much smaller than the repetition period, only the first term of the sum in \eq{ch6_equation14} is significant. 

In general, SPADs have negligible readout noise and the dark count rate contributes to the background noise. Thus, we can model the distribution of photons detected for each data point by a Poisson distribution of expectation $f_i$. The PDF associated with the observation of $X$ events on a given data point is then expressed by
\begin{equation}
p_i(X; \boldsymbol{\theta}) = \frac{f_i^{X}}{X!}e^{-f_i} \; .
\label{ch6_equation25}
\end{equation}
The set of parameters that must be estimated from the data is $\boldsymbol{\theta}=(N,\Gamma)$, while we estimate $q_{irf}(t)$, $q_b(t)$ and $N_b$ with independent measurements. Then, the information matrix can be calculated from \eq{ch6_equation4}.

\paragraph{Cramér-Rao bound}

After having experimentally measured the value of the parameters involved in the model, we can compute the Cramér-Rao bound on the standard error $\sigma_\Gamma$ on the decay rate estimates. The Cramér-Rao inequality can be expressed as
\begin{equation}
\frac{\sigma_{\Gamma}}{\Gamma} \geq \frac{1}{\sqrt{N}} \times F \left(T,N_b,q_{irf},q_{b},n \right) \; ,
\label{eq_ch6_0}
\end{equation}
where $n$ is the number of data points and $F$ is calculated by numerically inverting the information matrix \cite{bouchet_fisher_2018_2}.

\bibliographystyle{apsrev_no_url}

%\bibliography{references}

\end{document}